\documentclass[useAMS,usenatbib]{mn2e}
\usepackage{graphicx}
\usepackage{amssymb}

\def\deg{\ifmmode^\circ\else$^\circ$\fi}

\def\msun{M$_{\odot}$}

\newcommand{\mic}{\,$\mu$m}
\newcommand{\lmst}{$\ell_{\rm MST}$}
\def\Q{\ifmmode\mathcal{Q}\else$\mathcal{Q}$\fi}
\def\Mach{\ifmmode\mathcal{M}\else$\mathcal{M}$\fi}

\title[The structures of embedded clusters in nearby molecular clouds]
{The structures of embedded clusters in the Perseus, Serpens and Ophiuchus molecular clouds}
\author[S. Schmeja, M. S. N. Kumar and B. Ferreira]
{S. Schmeja$^{1,2}$\thanks{E-mail: sschmeja@ita.uni-heidelberg.de},
M. S. N. Kumar$^{1}$ and B. Ferreira$^{3}$
\\
$^1$Centro de Astrof{\'{\i}}sica da Universidade do Porto,
Rua das Estrelas, 4150-762 Porto, Portugal \\
$^2$Zentrum f\"ur Astronomie der Universit\"at Heidelberg, 
Institut f\"ur Theoretische Astrophysik, Albert-Ueberle-Str.~2, 
69120 Heidelberg, Germany\\
$^3$Department of Astronomy, University of Florida, Gainesville, FL 32611-2055, USA \\
}
\begin{document}

\date{ }

\pagerange{\pageref{firstpage}--\pageref{lastpage}} \pubyear{2007}

\maketitle

\label{firstpage}

\begin{abstract} The young stellar population data of the Perseus, Ophiuchus and Serpens molecular 
clouds are obtained from the {\it Spitzer} c2d legacy survey in order to investigate the spatial 
structure of embedded clusters using the nearest neighbour and minimum spanning 
tree method.
We identify the embedded clusters in these clouds as density enhancements and
analyse the clustering parameter \Q\ with respect to source luminosity and evolutionary stage.
This analysis shows that the older Class~2/3 objects are more centrally condensed than the younger
Class~0/1 protostars, indicating that clusters evolve from an initial hierarchical 
configuration to a centrally condensed one.
Only IC~348 and the Serpens core, the older clusters in the sample, shows signs of mass segregation 
(indicated by the dependence of \Q\ on the source magnitude), pointing to a 
significant effect of dynamical interactions after a few Myr.
The structure of a cluster may also be linked to the turbulent energy in the natal cloud
as the most centrally condensed cluster is found in the cloud with the lowest Mach number
and vice versa.
In general these results agree well with theoretical scenarios of star cluster formation
by gravoturbulent fragmentation.
\end{abstract}


\section{Introduction}
\label{sec:intro}

It is believed that stars form from high-density regions in turbulent molecular clouds
that become gravitationally unstable, fragment and collapse.
That way, a star cluster is built up in a complex interplay between gravity and 
supersonic turbulence  \citep{maclow+klessen04, bp07, mckee+ostriker07}.
Therefore the structure of an embedded cluster, i.e.\ the spatial distribution of its
members, may hold important clues about the formation mechanism and initial conditions.
Several characteristics of star-forming clusters seem to be correlated with the
turbulent flow velocity, e.g.\ the star formation 
efficiency \citep*{klessen00,clark+bonnell04,skf05}, mass accretion rates \citep{sk04}, 
or the core mass distribution \citep{bp06} and the
initial mass function \citep{padoan+nordlund05}.

Giant molecular cloud (GMC) complexes usually contain multiple embedded
clusters as well as more distributed, rather isolated young stellar objects (YSOs).
The interstellar medium shows a hierarchical structure (sometimes described as fractal)
from the largest GMC scales down to individual cores and clusters, which are sometimes
hierarchical themselves. There is no obvious change in morphology at the cluster boundaries,
so clusters can be seen as the bottom parts of the hierarchy, where
stars have had the chance to mix \citep{efremov+elmegreen98, elmegreen00, elmegreen06}.

Since embedded stars are best visible in the infrared part
of the spectrum, and since nearby GMCs can span several degrees on the sky,
wide-field near-infrared (NIR) and mid-infrared (MIR) surveys
are useful tools for identifying and mapping the distribution of young
stars in GMCs.
The advent of the {\it Spitzer Space Telescope} has in particular led 
to large advances in such studies.
The {\it Spitzer} Cores to Disks (c2d) Legacy programme
survey \citep{evans03} aimed at mapping five large nearby star-forming
regions, including Perseus, Serpens and Ophiuchus
using {\it Spitzer's} IRAC (3.6 to 8\mic) and MIPS (24 to 160\mic) instruments
\citep{joergensen06, harvey06, harvey07a, harvey07b, rebull07,padgett07}.

The Perseus molecular cloud complex is an extended region of low-mass star formation
containing the well-studied embedded clusters IC~348 and NGC~1333.
The latter is considered to be significantly younger \citep*[$< 1$\,Myr;][]{lada96, wilking04}
than IC~348 \citep[$\sim 3$\,Myr;][]{luhman03,muench07}.
The distance to the Perseus cloud is uncertain and may lie somewhere between 200 and 320\,pc,
presumably there is a distance gradient along the cloud \citep[see e.g.\ the
discussion in][]{enoch06}.
Areas of 3.86\,deg$^2$ and 10.6\,deg$^2$ have been mapped by IRAC and MIPS, respectively
\citep{joergensen06, rebull07}.
\cite{joergensen06} identified a total number of 400 YSOs, thereof 158 in IC~348 and 98 in NGC~1333,
which means that a significant fraction of young stars is found outside the main
clusters \citep[see also][]{hatchell05}.
\cite{muench07}, also based on {\em Spitzer} MIR data, report a known membership
of IC~348 of 363 sources and estimate a total population of more than 400.

The Serpens molecular cloud is a very active star-forming region covering
more than 10\,deg$^2$ on the sky at a distance of about 260\,pc \citep*{straizys96}.
Areas of 0.89\,deg$^2$ and 1.5\,deg$^2$ have been mapped by IRAC and MIPS, respectively
\citep{harvey06, harvey07a}.
In this region, 235 YSOs have been identified by \cite{harvey07b}.
The age of the main cluster, called the Serpens cloud core, is estimated to be
$\sim 2$\,Myr \citep{kaas04}.

The Ophiuchus (or $\rho$~Ophiuchi) molecular cloud 
is one of the closest star-forming regions at
a distance of about 135\,pc \citep{mamajek07}.
Its main cluster, L1688, is the richest known nearby embedded cluster.
It is very young, with an age of probably $< 1$\,Myr \citep{greene+meyer95, luhman+rieke99}.
14.4\,deg$^2$ have been observed with MIPS, leading to the identification of 323 YSO
candidates \citep{padgett07}. 

In this paper we analyse the structure and distribution of the young stellar
population unveiled by the c2d programme in these nearby molecular clouds and compare
the results to current theoretical scenarios of star formation.
For this purpose, we retrieved the c2d point source catalogues of the Perseus, 
Serpens and Ophiuchus molecular clouds
and applied two complementary statistical methods to identify embedded clusters and
analyse their structures.
The selection of the data set is described in Section~\ref{sec:data}, the statistical methods
and their application to the data are explained in Section~\ref{sec:methods} and the results
and discussion are presented in Section~\ref{sec:analysis} and \ref{sec:discussion}.

\section{Data selection}
\label{sec:data}

The third delivery of data from the c2d legacy project \citep{evans05}, which combines
observations from the IRAC and MIPS cameras was obtained from the c2d
website\footnote{\tt http://ssc.spitzer.caltech.edu/legacy/c2dhistory.html}.
The combined photometric catalogues for the Perseus, Ophiuchus and
Serpens molecular clouds were retrieved. 
These catalogues are products resulting from extensive analysis of the photometric
observations in the four IRAC bands (central wavelengths 3.6, 4.5, 5.8 and 8.0\mic)
and the three MIPS bands covering the 24 to 160\mic\ wavelength range.
The sensitivity of the survey is expected to sample objects down to 0.1 - 0.5 \msun\
in all three clouds.

The delivery catalogues provide source identifications and classifications
based on a consistent method of analysis for all the targets covered by the 
legacy survey as described by \cite{evans05}.
The catalogues contain positions, photometry and associated quality flags,
source type such as point/extended source and source classification such as
stars/galaxy/YSO.
These catalogues focus on identifying and classifying YSOs and distinguishing them from
stars and galaxies. 
Extensive analysis of the point sources using colour-colour and colour-magnitude diagrams,
spectral indices in the observed bands, comparison with SWIRE ({\em Spitzer} Wide-area Infrared 
Extragalactic survey) data to remove
galaxy and PAH emission contaminants is used for this purpose.

\begin{table}
\centering
\caption{Classification scheme}
\label{tab:classification}
\begin{tabular}{l l}
\hline
object type & classification,     \\
\citep{evans05} & main criteria \\
\hline
red & {\bf Class 0/1}  \\
red1 &  (objects only visible at \\
red2 &  wavelengths $\ge 8$\,\mic)  \\
YSOc\_MP1\_red &  \\
YSOc\_IR4+MP1\_red &  \\
\hline
YSOc\_IR4 & {\bf Class 2/3}  \\
YSOc\_MP1 & (objects visible at \\
YSOc\_IR4+MP1 &  wavelengths $< 8$\,\mic,  \\
YSOc\_IR4\_PAH-em & SEDs with slopes \\
YSOc\_IR4\_star+dust(BAND) & typical for Class 2/3)  \\
YSOc\_IR4\_red &  \\
YSOc\_MP1\_PAH-em &  \\
YSOc\_MP1\_star+dust(BAND) &  \\
YSOc\_IR4+MP1\_PAH-em &  \\
YSOc\_IR4+MP1\_star+dust(BAND) &  \\
\hline
star & {\bf stars} \\
star+dust(BAND) & (SEDs fitted with \\
 & photospheres)\\
\hline
\end{tabular}
\end{table}

In the present work we use these catalogues to retrieve the stellar and YSO population
in the three molecular clouds based on the source classification flags.
Although YSOs are classified based on their spectral energy distributions 
(SEDs) and MIR colours, a certain amount of
confusion is always present when the complete SEDs are not available and only a part
of the SED is sampled.
The IRAC and MIPS observations cover the wavelength range where the YSOs lend themselves
for an effective classification, however, confusion arising due to partial sampling of
the SED cannot be removed. Further, an edge-on Class~2 source can imitate a Class~1 source
etc.\ when classification is made using partial SEDs. 
Based on the nature of the point source SED, which
is represented by 19 different flags in the c2d catalogues, we group the point sources
into three categories, namely stars (foreground/background and cloud members), YSOs in Class~0/1 phase
and YSOs in Class~2/3 phase.
While the c2d catalogue provides a distinct classification for stars, the YSOs are represented
by 15 different object types.
We use the thumb rule of grouping sources that are visible only in wavelengths longer
than 8\mic\ into the Class~0/1 group and the remaining into the Class~2/3 group.
This leads to the classification given in Table~\ref{tab:classification}, where
we relate the object types from \cite{evans05} and their characteristics
to the common evolutionary classes.
Our classification does not necessary represent the `true' nature of the objects,
which would in some cases require data from wavelengths longer than provided by {\em Spitzer}
and a more sophisticated SED modelling.

\begin{figure*}
\includegraphics[width=\textwidth]{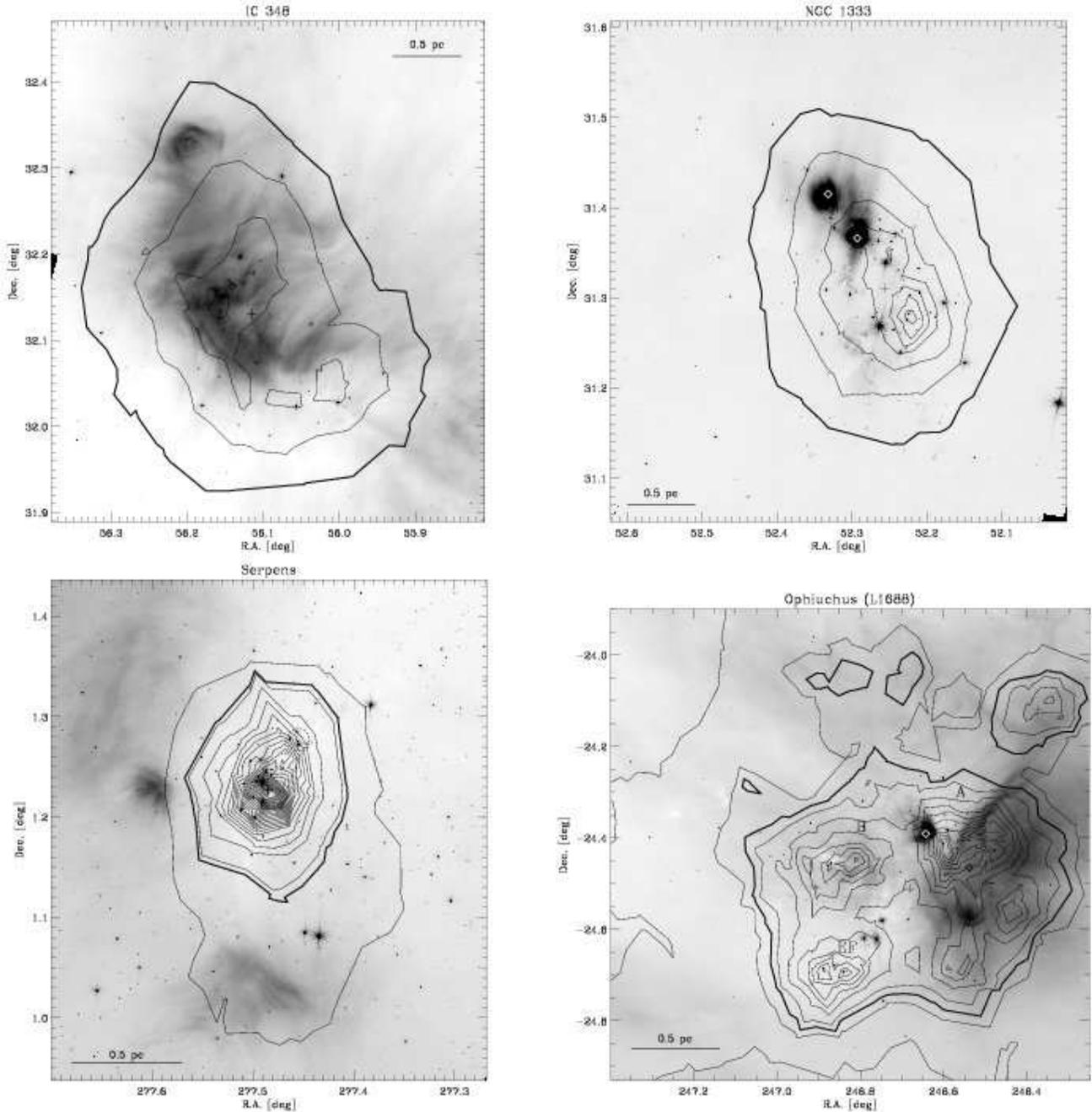}
\caption{{\it Spitzer} IRAC Channel~4 (8\mic) images of the four main clusters
IC~348, NGC~1333, Serpens~A and L1688
overlaid with the 20th nearest neighbour density contours.
The thin contours are plotted in intervals of 40\,pc$^{-2}$.
The outermost thin contours correspond to a nearest neighbour density of 
40\,pc$^{-2}$.
The thick contours indicate the cluster boundaries as defined in the text.
The horizontal bars showing a projected length of 0.5\,pc are based on assumed distances
of 315, 260 and 135\,pc to the Perseus, Serpens and Ophiuchus cloud, respectively.
Crosses indicate the cluster centres, the known B~stars are highlighted by diamond symbols.
}
\label{fig:clusters}
\end{figure*}

\begin{table*}
\centering
\caption{ Clustering parameters of the identified clusters in the three regions.}
\label{tab:embclusters}
\begin{tabular}{l c c c r r r c}
\hline
Region/ & RA (J2000) & Dec (J2000)  & size & $\rho_{\rm thresh}$ & $\rho_{\rm peak}$ & $n_*$ & \Q \\
\hspace{0.25cm} Cluster &  [deg] & [deg] & [pc] & [pc$^{-2}$] & [pc$^{-2}$] & & \\
\hline
{\bf Perseus}     &         &            &      &    &      &     &       \\
NGC 1333          &  52.2563&  +31.3107  & 2.09 & 20 &  261 & 158 & 0.81  \\
IC 348            &  56.1151&  +32.1310  & 2.53 & 20 &  140 & 204 & 0.80  \\
Barnard 3         &  54.9612&  +31.9227  & 1.67 & 20 &   45 &  51 & 0.69  \\
\hline
{\bf Serpens}     &         &            &      &    &      &    &      \\
Serpens core (A)  & 277.485 &   +1.2274 & 1.00 & 70 & 1045 & 84 & 0.84 \\
Serpens B         & 277.256 &   +0.5059 & 0.77 & 70 &  378 & 43 & 0.77 \\
\hline
{\bf Ophiuchus}   &         &            &      &    &      &     &       \\
Ophiuchus north   & 246.348 & $-$23.4660 & 0.78 & 95 &  126 &  34 & 0.68  \\
Ophiuchus centre  & 246.374 & $-$24.1197 & 0.52 & 95 &  213 &  29 & 0.73  \\
L1688             & 246.683 & $-$24.5112 & 1.85 & 95 &  509 & 337 & 0.72  \\
L1689S            & 248.045 & $-$24.9294 & 1.41 & 50 &  240 &  78 & 0.78  \\
\hline
\end{tabular}
\end{table*}

\section{Statistical Methods}
\label{sec:methods}

\subsection{Nearest neighbour method}

While clusters are readily identifiable as peaks in stellar density maps,
the exact delimitation of a cluster is difficult and will always be somewhat arbitrary,
since all the embedded clusters are surrounded by an extended population of YSOs
distributed throughout the entire clouds.
Several methods to detect clusters in a field have been developed.
Usually they are based on stellar densities (derived e.g.\ from star counts or
Nyquist binning) and consider as clusters all regions above
a certain deviation ($\sim 2-4 \sigma$) from the background level
\citep*[e.g.][]{lada91,lada+lada95,kumar06,froebrich07}.

The nearest neighbour (NN) method \citep{vonhoerner63, casertano+hut85}
estimates the local source density $\rho_j$ by measuring 
the distance from each 
object to its $j$th nearest neighbour, where the value of $j$ is chosen 
depending on the sample size. 
There is a connection between the chosen $j$~value and the sensitivity to those density
fluctuations being mapped.
A small $j$~value increases the locality 
of the density measurements at the same time as increasing sensitivity to random density 
fluctuations, whereas a large $j$ value will reduce that sensitivity at the cost of losing 
some locality.
Through the use of Monte Carlo simulations Ferreira \& Lada (in preparation)
find that a value of $j = 20$ is adequate to detect clusters with 10 to 1500 members.

The positions of the cluster centres are defined as the density-weighted enhancement centres 
\citep{casertano+hut85}
\[
x_{d,j} = \frac{\sum_i x_i \rho_j^i}{\sum_i \rho_j^i},
\]
where $x_i$ is the position vector of the $i$th cluster member and $\rho_j^i$
the $j$th NN density around this object.
These centres do not necessarily correspond to the density peaks.

\subsection{Minimum spanning tree and \Q}

The second method makes use of a minimum spanning tree (MST), 
a construct from graph theory, which is defined as
the unique set of straight lines (``edges'') connecting a given set
of points without closed loops, such that the sum of the edge
lengths is a minimum \citep{kruskal56, prim57}.
From the MST we derive the mean edge length \lmst.
\citet{cw04} introduced the parameter $\Q = \bar{\ell}_{\rm MST}/\bar{s}$, 
which combines the normalized correlation length $\bar{s}$,
i.e.\ the mean distance between all stars, and the normalized mean edge 
length $\bar{\ell}_{\rm MST}$.
The \Q\ parameter permits to quantify the structure of a cluster and to 
distinguish between clusters with a central density concentration and 
hierarchical clusters with possible fractal substructure.
Large \Q\ values ($\Q > 0.8$) describe centrally condensed clusters
having a volume density $n(r) \propto r^{-\alpha}$, while small \Q\ values ($\Q < 0.8$) 
indicate clusters with fractal substructure.
\Q\ is correlated with the radial density exponent $\alpha$ for $\Q > 0.8$ and anticorrelated with 
the fractal dimension $D$ for $\Q < 0.8$.
The dimensionless measure \Q\ is independent of the number of objects and
of the cluster area. 
The method has been successfully applied to both observed clusters and results of numerical
simulations \citep{cw04, sk06, ks07}.
For details of the method, in particular its implementation and normalization used for
this study, see \cite{sk06}.

\subsection{Application to the data}

The NN and MST statistical methods described above were applied to the c2d data
of the Perseus, Serpens and Ophiuchus molecular clouds.
We first used the NN analysis (described in detail by Ferreira et al., in prep.)
on the stellar and YSO populations in the entire clouds
to detect density enhancements and therefore embedded clusters.
For this purpose, 20th NN density maps were generated for the full extent of the available c2d data.
The 20th NN density is thought to be the optimum
density to identify primary clusters and not pick up very loose groups 
(Ferreira \& Lada, in preparation).
We found that using only the point sources that were classified as YSOs to produce
the NN maps yields the best density contrast and boundary identifications.
We determined the average 20th NN density in the clouds away from the clusters
and define as cluster members all YSOs whose density exceeds $3 \sigma$ of the
average value.
The cluster boundaries are shown as thick lines in Fig.~\ref{fig:clusters},
the density thresholds $\rho_{\rm thresh}$ for the clusters are given in
Table~\ref{tab:embclusters}.

All YSOs enclosed within the respective cluster defining contour are
treated as cluster members.
For those we constructed the MST and determined \Q.

Note that the exact definition of the cluster boundaries is only relevant for
the quantitative structure analysis using the \Q\ parameter.
Changing the cluster definition criteria may change the absolute \Q\ values, but it does not affect
the general trends and correlations discussed below.

\section{Structure Analysis}
\label{sec:analysis}

\subsection{Cluster morphology}

The YSOs in the three investigated molecular clouds are found in the embedded clusters and in a
distributed population that extends through the entire cloud
\citep{joergensen06,harvey07b}.
Figure~\ref{fig:clusters} shows IRAC Channel~4 (8 \mic) images of
the four largest embedded clusters IC~348, NGC~1333, Serpens cloud core and L1688,
overlaid with the 20th NN density contours.

In the Perseus molecular cloud, apart from the embedded clusters NGC~1333 and IC~348,
the region around Barnard~3 (B3) emerges as a moderately sized cluster, but no 
significant density enhancements are found associated with L1455 and L1448.
Despite the large number of member stars, the Perseus clusters show relatively small
peak densities compared to the other clouds.
Likewise, the Perseus cloud shows the lowest ``background'' YSO density of all regions.
The density peak in IC~348 coincides with the B5 star HD~281159, the most massive star
in this cluster, whereas the two B~stars in NGC~1333 (BD~+30~549 and SVS~3)
are not associated with the densest region.
The B0~star HD~278942 is found close to the centre of the B3 cluster, however, it may
be located behind the main Perseus cloud complex \citep{ridge06}.

In the Serpens cloud two major clusters were detected: The main cluster in this cloud, 
usually called the Serpens cloud core, and referred to as Cluster~A by \cite{harvey06}, 
and a recently discovered cluster called Cluster~B by \cite{harvey06}
and Serpens/G3-G6 by \cite{djupvik06}.
The Serpens core has a very steep density gradient and a prominent peak with a density
of 1045\,pc$^{-2}$, the highest of all investigated clusters and twice the value of the
second densest cluster.

In the Ophiuchus cloud one highly structured, major embedded cluster was detected
roughly centred on L1688, several smaller density enhancements clearly trace four filamentary
structures that seem to converge on this prominent cluster.
These density enhancements are in good correlation with the dust continuum emission maps.
Two of these density enhancements are picked up as small clusters by the NN method, we
designate them Ophiuchus north and centre.
The Ophiuchus north cluster is centred around a group of bright B~stars ($\rho$~Oph~D, $\rho$~Oph~AB, $\rho$~Oph~C)\footnote{These designations are not to be confused with the ones used for the subclusters
discussed below and marked in Fig.~\ref{fig:clusters}.}.
Furthermore we detect the known cluster associated with L1689S \citep{bontemps01},
which shows a single peak and no distinct substructure.

The shapes of IC~348, NGC~1333, Serpens~A and L1689S can be roughly approximated by ellipses,
whereas the hierarchical L1688 cluster shows an irregular, nevertheless rather circular, perimeter.
It is intersting to note that the elliptical clusters appear to have 
two peaks roughly coinciding with the foci of ellipses.
However, these peaks seem to consist of of YSO populations of different
evolutionary stages.
This is seen in NGC~1333, with the density peak in the south and two luminous, massive
B~stars, which illuminate the prominent reflection nebula, in the north.
In Serpens we note a group of bright stars north of the density peak, marked by a dotted
circle in Fig.~\ref{fig:clusters}.
The second peak in IC~348, seen to the south-east,
seems to correspond to a new star formation episode close
to HH211 \citep*{tafalla06}.

Table~\ref{tab:embclusters} lists all the detected embedded clusters 
along with their position, size, peak density, the \Q\ parameter
and the number of YSOs $n_*$. 
The size corresponds to the length of the major axis of an ellipse
fitted to the cluster areas. 
Its only purpose is to provide the reader with a rough estimate of the diameter of the cluster.
The linear sizes are computed assuming distances
of 315, 260 and 135\,pc to the Perseus, Serpens and Ophiuchus cloud, respectively.

\begin{figure*}
\includegraphics[width=\textwidth]{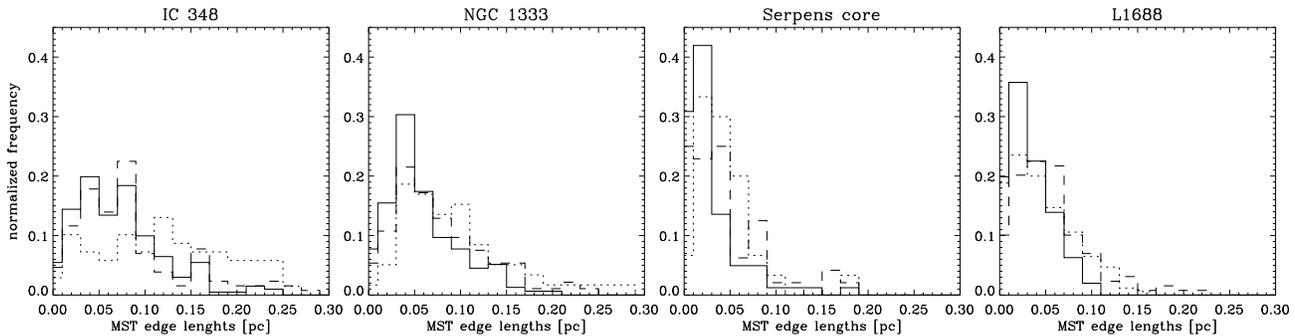}
\caption{Normalized histograms of MST edge lengths for Class 0/1 (dotted line), Class 2/3 (dashed line)
and all YSOs (solid line) in the four main clusters.} 
\label{fig:lmst_hist}
\end{figure*}

\subsection{\Q\ values}
\label{sec:Q}

Extensive analysis of the distribution of the MST edge lengths as a function of
YSO class and the \Q\ parameter as a function of YSO brightness was carried out only
for four embedded clusters where the number statistics are significant
and suitable for this kind of analysis. 
These four major embedded clusters are NGC~1333, IC~348, Serpens core and L1688.
It can be seen from Figure~\ref{fig:clusters} that the Serpens core, NGC~1333 and IC~348 
clusters are centrally concentrated, 
although at very different surface densities, whereas L1688 shows
at least four peaks enclosed within the cluster defining contour.
This is reflected in the \Q\ values listed in Table~\ref{tab:embclusters}.
L1688 shows the lowest \Q\ value (0.72) indicating fractal substructure,
whereas the other clusters show \Q\ values of $\Q \ge 0.8$, denoting central
condensation.
All the small clusters have values of $\Q < 0.8$.

%

In Table~\ref{tab:classes} we list the absolute and relative numbers of Class~0/1 
and Class~2/3 objects and their \Q\ values in the four large embedded clusters.
Note that the classifications (see Sect.~\ref{sec:data}) do not necessarily
correspond to the standard definitions.
Assuming that the ratio of younger to older sources represents 
the relative ages of the embedded clusters, the data in Table~\ref{tab:classes} indicate that the L1688 cluster
is the youngest of all, NGC~1333 and Serpens are slightly older and IC~348 is the oldest of all four.
This is not identical, but quite similar to the age estimates based on more sophisticated methods
mentioned in Sect.~\ref{sec:intro}.
 In all the clusters but Serpens, the \Q\ values
are significantly higher for the Class 2/3 objects than for the Class
0/1 protostars, indicating a more centrally condensed configuration
for the older sources.
In the centrally condensed Serpens A cluster the two groups
show the same \Q\ values.

\begin{table}
\centering
\caption{Numbers of objects and \Q\ values for the different evolutionary classes }
\label{tab:classes}
\begin{tabular}{l r r r r r r r}
\hline
Cluster & \multicolumn{2}{c}{Class 0/1} & \multicolumn{2}{c}{Class 2/3}  \\
& \multicolumn{1}{c}{$n_*$} & \multicolumn{1}{c}{\Q} & \multicolumn{1}{c}{$n_*$} & \multicolumn{1}{c}{\Q}        \\
\hline
NGC 1333  &  62 (39\%) & 0.78 &  96 (61\%) & 0.83 \\
IC 348    &  72 (35\%) & 0.71 & 132 (65\%) & 0.78 \\
Serpens A &  33 (38\%) & 0.85 &  51 (62\%) & 0.85 \\
L1688     & 194 (58\%) & 0.69 & 143 (42\%) & 0.78 \\
\hline
\end{tabular}
\end{table}

\begin{table}
\centering
\caption{Clustering parameters of the subclusters in L1688}
\label{tab:l1688_subclusters}
\begin{tabular}{l c c c c c}
\hline
Subcluster & RA         & Dec     & $\rho_{\rm peak}$ & $n_*$ & \Q \\
           &  [deg]     & [deg]   & [pc$^{-2}$]       &       &     \\
\hline
Oph A     & 246.553  &  $-$24.416 & 509 & 67 & 0.74 \\
Oph B     & 246.836  &  $-$24.465 & 373 & 52 & 0.79 \\
Oph EF    & 246.854  &  $-$24.679 & 367 & 43 & 0.77 \\
\hline
\end{tabular}
\end{table}

The three major subclusters in L1688 were investigated further, these are (following the
nomenclature of \citealt{bontemps01}) Oph~A, Oph~B and Oph~EF (marked in Fig.~\ref{fig:clusters}).
They are defined as having NN densities $> 180$\,pc$^{-2}$, their clustering properties
are listed in Table~\ref{tab:l1688_subclusters}.
Oph~A contains 41 Class~0/1 protostars and 26 Class~2/3 objects and has a value of
 $\Q = 0.74$ in the hierarchical range.
Oph~B (18 Class~0/1, 34 Class~2/3 sources) and Oph~EF (16 Class~0/1, 27 Class~2/3 sources)
show higher values of $\Q = 0.79$ and $Q = 0.77$, respectively.
Oph~A is associated with a larger amount of gas and has a significantly higher
Class~0/1 to Class~2/3 ratio than the other two subclusters, making it most likely the
youngest of the subclusters, it shows the lowest \Q\ value of the subclusters.
Again we see an evolutionary sequence from the youngest, hierarchical cluster to the more
evolved clusters closer to central condensation.
When looking at Class~0/1 and Class~2/3 sources separately, we again find significantly
higher \Q\ values for the more evolved objects: In Oph~A the value rises from $\Q = 0.74$ for
Class~0/1 to $\Q = 0.90$ for Class~2/3, in Oph~B from 0.77 to 0.81 and in Oph~EF from 
0.73 to 0.79.

\subsection{Distribution of MST edge lengths}

The lengths of the MST edges can be seen as the separation from one object to its
nearest neighbour.
Histograms of the MST edge lengths indicate characteristic separations for YSOs in each
of the clusters. 
Figure~\ref{fig:lmst_hist} shows the normalized histograms of the MST edge lengths
in the four main clusters in bins of 0.02\,pc, using the assumed distances given above 
(Section~\ref{sec:Q}).
The histograms show a distinct peak at 0.02\,pc (Serpens, L1688) and 0.04\,pc (NGC~1333),
respectively, and the distributions spread up to separations of 0.2 to 0.3\,pc.
In IC~348, however, two peaks appear at a larger separation of 0.04 and 0.08\,pc.
Also shown in Fig.~\ref{fig:lmst_hist} are the same histograms for Class~0/1 and Class~2/3
sources separately.
In the clusters NGC~1333, Serpens and L1688 we do not see a significant difference between
the separations of Class~0/1 and Class~2/3 objects.
Objects of all evolutionary states are more or less similarly distributed in the cluster.
In IC~348, however, it is only the Class~2/3 objects which roughly follow the distribution
of the overall cluster members, whereas the Class~0/1 objects seem to be homogeneously 
distributed.
This is because the central cluster in IC~348 is evolved and has blown out a cavity,
with the current star formation occurring in a shell around the cluster \citep{tafalla06, muench07}.

The characteristic edge lengths found from Fig.~\ref{fig:lmst_hist} do reflect the
true separations and are not a result of poor spatial sampling  because these values 
are almost an order of magnitude higher than the tolerance used for matching
IRAC and MIPS data (4 arcsec, corresponding to 0.003 - 0.006\,pc).
We also investigated the MSTs and their edge lengths for the stars in the cluster field,
but due to their large spatial density they would generally yield very short edges and a
narrow peak of the distribution at very small projected separations.

\begin{figure*}
\includegraphics[width=14cm]{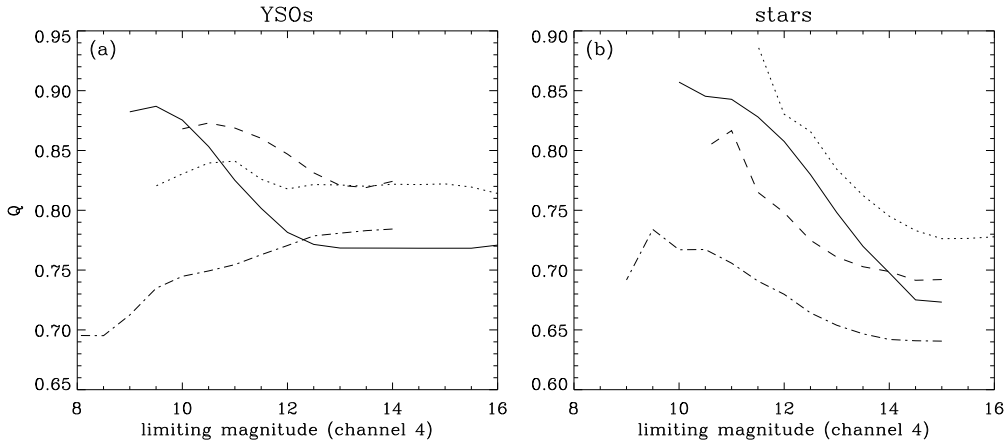}
\caption{\Q\ as a function of source magnitude for YSOs (left) and
stars (right) in Channel~4 for IC~348 (solid line), NGC~1333 (dotted line), Serpens~A (dashed line) and
L1688 (dash-dotted line).} 
\label{fig:Q_vs_mag}
\end{figure*}

\subsection{\Q\ versus magnitude}

Assuming that the IR source magnitudes are representative of the source masses
and evolutionary stages we investigate the variation of \Q\
as a function of magnitude.
\Q\ was calculated for sources brighter than a certain magnitude
in steps of 0.5\,mag as long as the sample contained at least 40 objects.
This calculation was made for YSOs and stars separately. 
The resulting \Q\ versus Channel~4 magnitude curves for YSOs and stars are shown in 
Fig.~\ref{fig:Q_vs_mag}a and b respectively. 
(Similar plots were made using Channel~1, 2 and 3 magnitudes as well; they look
similar in all bands.)
The results for different clusters are shown using different line styles.
Fig.~\ref{fig:Q_vs_mag}a shows that the YSO curves remain flat or rising with magnitude
for the clusters NGC~1333 and L1688 while they decrease with magnitude for
IC~348 and, less pronounced, for Serpens.
This implies that in NGC~1333 and L1688 YSOs of all luminosities
display a roughly similar configuration, whereas in Serpens and IC~348 the low luminosity YSOs
are less centrally concentrated than the more luminous YSOs.

In contrast, the stellar population shows a declining \Q\ versus magnitude curve for
all the clusters, indicating that the fainter stars are relatively homogeneously
distributed in space and the brighter stars are centrally concentrated. Since this curve 
is computed for a region encompassing only the dense cluster, the chances of picking up 
fainter background sources through the large column of extinction is low. 
Due to the same reason, stars that are cluster members outnumber foreground stars for 
the computed area. Therefore the observed effect is representative of the true variations
among the cluster members. Furthermore,
a random distribution expected from foreground or background objects will produce a \Q\ value 
constant over the whole magnitude range at $\Q \approx 0.73$ and cannot reproduce the declining curve
seen in Fig.~\ref{fig:Q_vs_mag}b.

Assuming the age of IC~348 as 3\,Myr, that of Serpens as 2\,Myr and that of NGC~1333 
and L1688 around 1\,Myr (see Sect.~\ref{sec:intro}), the YSO curves in Fig.~\ref{fig:Q_vs_mag}a
show that the brighter 
objects appear to evolve towards centrally condensed configurations and fainter objects 
towards homogeneous distributions with time.
The transition of the distribution of the lowest-mass/substellar objects from a centrally
condensed configuration to a more homogeneous distribution is thought to be an
effect of dynamical evolution.
A detailed discussion of this effect in IC~348 and the Orion Trapezium cluster and its implications
for the formation of brown dwarfs can be found in \cite{ks07}.

\section{Discussion}
\label{sec:discussion}

\subsection{Dynamical evolution and mass segregation}

The structure analysis of YSOs and stars presented in the previous section has several
implications.
The significantly higher \Q\ values for the Class 2/3 objects compared to the Class 0/1 protostars
show that the older sources a more centrally condensed than the younger objects, which 
in most cases show a hierarchical configuration.
Also in other embedded clusters the Class 0/1 protostars are found in subclusters,
e.g.\ in NGC~2264 \citep{teixeira06} or the Orion Trapezium cluster \citep{lada04}.
This is in agreement with the predictions of numerical simulations \citep*{bbv03,sk06}
suggesting that clusters build up from several subclusters that eventually
form a single centrally condensed cluster.
However, it is not possible to distinguish between the structure inherent to the cloud and that 
arising due to dynamical evolution.
It is tempting to assume that the Class~0/1 objects trace the structure inherent to
the cloud while the older objects represent the structure as a result from dynamical
evolution.
More than half of the mass of a Class~0 protostar is found in its envelope.
So a Class~0 source can be seen as still part of the cloud, while 
the older YSOs are star-like objects with a disc rather than an envelope
and are therefore susceptive to dynamical interaction.

While \Q\ rises as a cluster evolves, its absolute value does not depend 
on the age of the cluster. 
NGC~1333 and IC~348 with estimated ages of 1 and 3 Myr, respectively, have similar \Q\ values
of 0.81 and 0.80, whereas the two extreme \Q\ values in our sample (0.84 and 0.72) are 
associated with the clusters L1688 and Serpens thought to have ages of about 1 and 2 Myr,
respectively.
This shows that the cluster structure in general is not directly correlated with the evolutionary stage
and rather depends on the existing large-scale structure of the cloud, but the observation that Class~2/3 objects (which had more time and are also more susceptive to dynamical interactions) are more centrally condensed than the younger protostars 
is a clue that the cluster evolves from a hierarchical to a more centrally condensed configuration as predicted by numerical simulations.

We interpret the decrease of \Q\ with fainter (i.e.\ less massive) sources as an effect
of mass segregation.
However, the relation of luminosity to mass may be different for the different
evolutionary classes, which may cause some interference of evolutionary and mass segregation effects.
The trend shown in Fig.~\ref{fig:Q_vs_mag}a, based on Channel~4 (8\,\mic) magnitudes,
is mainly representative of Class~2/3 objects, but it is similar in the other IRAC bands,
and also in MIPS Channel~1 (24\,\mic), which includes Class~0/1 objects.

A decrease of \Q\ with fainter magnitudes is seen in IC~348 ($\sim 3$\,Myr) and,
 to a somewhat lesser extent, in 
Serpens ($\sim 2$\,Myr), whereas there are no signs of mass segregation in the YSO 
population of the youngest clusters.
The stellar (i.e.\ older) population, on the other hand, shows significant mass segregation
in all clusters.
Therefore mass segregation seems to be a function of age.
According to simulations of turbulent molecular clouds \citep{bbv03} YSOs form at different 
locations in the cloud and condense into a single cluster
within a few free-fall times or less than 1 Myr.
Dynamical interactions become important only after this initial assembly of YSOs is made.
In particular, dynamical mass segregation occurs on a time-scale of the order
of the relaxation time of the cluster \citep{bonnell+davies98}.
This is consistent with the observed effects.
Our findings are not necessarily contradicting claims that mass segregation is primordial,
i.e.\ that massive stars form near or in the centre of a cluster \citep{bonnell+davies98}.
In the youngest clusters, contrary to the older ones, the massive 
YSOs show a more hierarchical distribution (smaller \Q, see Fig.~\ref{fig:Q_vs_mag}a)
than the lower-mass objects,
suggesting that the massive stars are still on the bottom of the potential wells of the
former subclusters.
This is consistent with theoretical scenarios in which mergers of small clumps that are 
initially mass segregated (or in which mass segregation is generated quickly by two-body 
relaxation) lead to larger clusters that are also mass segregated \citep*{mcmillan07}.

The NN maps (Fig.~\ref{fig:clusters}) indicate that IC~348, NGC~1333
and Serpens~A have formed from a single dense core, while
L1688 appears to lie at the interface of merging filaments (Ferreira
et al., in preparation).  L1688 has a highly hierarchical structure and
consists of at least three readily identifiable subclusters,
nevertheless it is detected as a single cluster by the 20th NN maps,
whose centre does not coincide with any of the subclusters. Provided
this geometrical centre corresponds approximately to the centre of
mass and provided the density is high enough, the subclusters may
merge to form a single centrally condensed cluster in the future.  The
structure of the cluster seems to be correlated to the cloud structure
in the sense that the collapse of a single dense core more
easily leads to a centrally condensed cluster, while interacting
filaments are reflected in a longer-lasting hierarchical structure of
the cluster.

\subsection{Turbulence and cluster structure}

\begin{table}
\centering
\begin{minipage}{80mm}
\caption{Gas properties and Mach numbers}
\label{tab:mach}
\begin{tabular}{l c c c c c l}
\hline
Cluster & $\Delta v$   & $T$ & $c_s$    &  \Mach & \Q & Ref.\footnote{References for $\Delta v$ and $T$: 1: \cite{jijina99}, 2: \cite{myers78} }  \\
 &  {\small[km s$^{-1}$]} & {\small [K] } & {\small [km s$^{-1}$] } &   &  & \\
\hline
Serpens A & 0.77         & 12  & 0.21        &  3.8 &  0.84 & 1  \\
NGC 1333  & 0.95         & 13  & 0.21        &  4.5 &  0.81 & 1  \\
L1688     & 1.33         & 15  & 0.23        &  5.8 &  0.72 & 2  \\
\hline
\end{tabular}
\end{minipage}
\end{table}

As turbulence seems to play an important r\^ole in the build-up of a star cluster
\citep{elmegreen00,maclow+klessen04,bp07},
we investigate the relation of the cluster structure with the turbulent environment.
In Table~\ref{tab:mach} we list the velocity dispersion $\Delta v$ and
temperature $T$ of the dense gas and the derived Mach numbers 
in Serpens, NGC~1333 and L1688 as traced by NH$_3$ line observations \citep*{myers78, jijina99}.
The NH$_3$ line is a tracer of dense gas which is representative of the cores from which 
star clusters are born \citep{bergin+tafalla07} and is also the tracer that it least affected by outflows.
Therefore, the NH$_3$ line widths better represent the turbulent properties of the 
cluster forming cores than other tracers of the molecular cloud such as CO or CS,
which may be seriously affected by the presence of outflows and effects of molecular
depletion \citep{tafalla02,bergin+tafalla07}.
Similarly, the best measure of the core temperatures is also obtained from the same emission
lines.
Hence we use the observed values of $\Delta v$ and $T$ from this tracer to compute the Mach number 
$\mathcal{M} = \Delta v/c_s$ where the isothermal sound 
speed $c_{\rm s} = (R T/\mu)^{1/2}$ with the specific gas constant $R$ and the mean molecular weight
$\mu = 2.33$.
\cite{myers91} argue that core linewidths are dominated by non-thermal motions 
(turbulence) only for core masses above $\sim 22$\,\msun\ and that linewidths are 
thermal below  $\sim 7$\,\msun. 
Since the core masses in all three clouds are above $\sim 30$\,\msun\ at an $A_V \approx 20$\,mag
contour \citep{enoch07}, the assumption that the linewidths are caused by turbulence is
justified.
The findings of \cite{myers91} also lead to the conclusion that cluster-forming cores are
turbulent while quiescent cores form isolated stars \citep[see also][]{myers01}.

Comparing the Mach numbers and \Q\ values in Table~\ref{tab:mach}
shows that Serpens, being the most centrally condensed cluster, is
associated with the lowest Mach number (turbulent energy) and L1688
being the most hierarchically structured cluster is associated with
the highest Mach number (see also Fig.~\ref{fig:clusters}).  Note that
we compare \Q\ and \Mach\ only for the three clusters that have a
significant population of YSOs, are embedded in dense gas and have
similar ages of 1 - 2\,Myr.  For this reason we excluded IC~348 from
this comparison, which is significantly older and largely depleted of
cold gas.  Since \Q\ seems to depend on the temporal evolution, a
comparison only makes sense for clusters of roughly the same age.

These results are in accordance with the scenario of turbulent fragmentation
and hierarchical formation of clusters.
Numerical simulations demonstrate that high Mach number flows lead to a highly fragmented
density structure \citep{bp06}.
\cite{enoch07} study the same clusters and obtain a contradictory result; however, 
the Mach numbers they compute are based on velocity dispersions obtained from observations of the
CO molecule, which rather traces the outer envelope of the molecular cloud and may
be contaminated by the effects of outflows.

As this correlation is obviously based on small-number statistics and has not been
found in numerical simulations of gravoturbulent fragmentation \citep{sk06} and considering
the associated uncertainties of $\Delta v$ and $T$, it has to remain rather
speculative.
Nevertheless it is a hint that the structure of an embedded cluster may be somehow connected to
the turbulent environment.
Even though the turbulent velocity is probably not the only agent responsible for shaping the cluster,
it may play a significant r\^ole.
While a high degree of turbulence in the cloud may keep the young stars in a more hierarchical distribution
for a longer time, the absence of strong turbulent motions may help to reach a
centrally condensed configuration more quickly.
Assuming that strong turbulence in the cloud also leads to a high velocity dispersion among the embedded
stars, this is also supported by the results of
\cite{goodwin+whitworth04}, who performed N-body simulations of the dynamics of fractal star clusters 
in order to investigate the evolution of substructure in recently formed clusters.
They show that if the velocity dispersion in the cluster is low, much of the substructure
will be erased, however if the velocity dispersion is high, significant levels of substructure
can survive for several crossing times.

\section{Summary}
\label{sec:summary}

We used the {\it Spitzer} c2d survey data of the nearby molecular clouds Perseus,
Serpens and Ophiuchus to identify embedded clusters and analyse their structures
using the nearest neighbour and MST method.
Apart from the large known embedded clusters IC~348, NGC~1333, Serpens core and
L1688 we only found a few relatively small clusters.
The \Q\ parameter was determined for all embedded clusters in these clouds and the MST edge lengths
and \Q\ values were analysed as a function of YSO class and magnitudes.
Our main results can be summarised as follows:

\begin{enumerate}

\item
Among the four large clusters, IC~348, NGC~1333 and Serpens are centrally condensed
($\Q \ge 0.8$), while L1688 shows a hierarchical structure with several density peaks.
The three main subclusters in L1688 show again a hierarchical structure, with
Oph~A, presumably the youngest one, having the lowest \Q\ value.
The peak densities vary strongly between the clusters in different clouds,
as does the density of the dispersed YSO population.

\item
In all clusters the \Q\ values for the younger Class~0/1 objects are substantially 
lower than for the more evolved Class~2/3 objects. This indicates
that embedded clusters are assembled from an initial hierarchical configuration
arising from the filamentary parental cloud and eventually end up as a single 
centrally condensed cluster, as predicted by numerical simulations of turbulent
fragmentation.
While the youngest, deeply embedded objects may represent the structure
inherent to the cloud, the distribution of the older objects probably is a 
result from dynamical interactions.

\item
We find no signs of mass segregation for the YSOs in the youngest clusters,
only for the YSO population in the presumably older clusters Serpens and IC~348.
In contrast, the stellar population displays a clear mass segregation in all
clusters.
This suggests that the effect of dynamical interaction becomes visible
at a cluster age between about 2 and 3\,Myr.

\item
The structure of a cluster may be related to the turbulent velocity in the
cloud in a way that clusters in regions with low Mach numbers reach a centrally condensed
configuration much faster than those in highly turbulent clouds,
in agreement with N-body simulations \citep{goodwin+whitworth04}.

\end{enumerate}

The results from the Perseus, Serpens and Ophiuchus star-forming regions 
are in good agreement with the predictions of theoretical scenarios
claiming that embedded clusters form from gravoturbulent fragmentation
of molecular clouds in a hierarchical process.
Our results may not be applicable to the YSO populations in other
regions, in particular the Taurus star-forming region, which shows a level
of clustering and velocity dispersions very different from the regions
studied in this work and which is generally harder to reconcile with the
gravoturbulent scenario \citep{fssk06}.

\section*{Acknowledgments}

This work is based on observations made with the Spitzer Space Telescope, which is 
operated by the Jet Propulsion Laboratory, California Institute of Technology under
a contract with NASA.
SS and MSNK were supported by a research grant POCTI/CFE-AST/55691/2004
approved by FCT and POCTI, with funds from the European community programme
FEDER.
SS also acknowledges funding by the Deutsche Forschungsgemeinschaft (grant SCHM~2490/1-1) 
during part of this work.
MSNK is also supported by the project PTDC/CTE-AST/65971/2006 approved by FCT.
BF wants to acknowledge support from his PhD advisor Elizabeth Lada as well
as financial support from NSF grant AST 02-02976 and National
Aeronautics and Space Administration grant NNG 05D66G issued to the
University of Florida.

\label{lastpage}

\end{document}